\documentclass[
twocolumn,
superscriptaddress,
pra,
longbibliography,
showpacs,
preprintnumbers,
amssymb,
floatfix
]{revtex4-1}

\usepackage{amsmath}
\usepackage{bm}
\usepackage{graphicx}
\usepackage[ansinew]{inputenc}
\usepackage{array}
\usepackage{color}
\usepackage{amsxtra}
\usepackage{amstext}
\usepackage{amsthm} 
\usepackage{amssymb}
\usepackage{latexsym}
\usepackage{gensymb}
\usepackage{dsfont}
\usepackage{braket}
\usepackage[caption=false]{subfig}

\usepackage[makeroom]{cancel}

\usepackage{balance}
\usepackage{dcolumn}
\usepackage{bm}
\usepackage{bbm}
\usepackage{braket}
\usepackage{mathrsfs}
\usepackage{euscript}
\usepackage{comment}
\usepackage{txfonts}

\usepackage[dvipsnames]{xcolor}

\usepackage[
unicode=true,
bookmarks=false,
breaklinks=false,
pdfborder={0 0 1},
backref=false,
colorlinks=true,
citecolor=blue
]{hyperref}

\usepackage{tikz-cd}

\usepackage{mathrsfs}

\usepackage{amsthm}

\DeclareMathOperator{\Tr}{Tr}

\begin{document}


\title{Optimal Quantum Illumination with Nonlocal Non-Gaussian Operations}

\author{Luis D. Zambrano Palma}
\affiliation{Institute for Quantum Science and Engineering, Texas A\&M University, College Station, Texas 77843, USA}

\author{Yusef Maleki}

 \affiliation{Institute for Quantum Science and Engineering, Texas A\&M University, College Station, Texas 77843, USA}

\author{M. Suhail Zubairy}

\affiliation{Institute for Quantum Science and Engineering, Texas A\&M University, College Station, Texas 77843, USA}

\date{\today}

\begin{abstract}
Enhancing quantum illumination with highly entangled probes remains an active area of research. In this context, non-Gaussian operations provide an effective route for engineering probe states that can surpass the standard two-mode squeezed state (TMSS).
In this work, we investigate a specific nonlocal non-Gaussian operation protocol and show that the engineered state using this protocol outperforms previously considered local non-Gaussian scenarios, engineered based on photon catalysis, addition, and subtraction under realistic conditions, including photon loss. Furthermore, by employing a $50{:}50$ beam splitter with photon-number difference detection, we demonstrate a significant enhancement in the signal-to-noise ratio (SNR) for target detection relative to the TMSS. Thus, our protocol exhibits improved performance, highlighting a resource-efficient and experimentally feasible probe for enhanced quantum illumination.
\end{abstract}

\maketitle

\section{Introduction}
The application of quantum-mechanical principles underlies many emerging technologies, including quantum cryptography~\cite{pirandola2020advances}, quantum computing~\cite{fitzsimons2017unconditionally,nielsen2010quantum}, and quantum metrology~\cite{pirandola2018advances}, all of which are fundamentally rooted in quantum processing and the efficient information extraction from physical systems and quantum measurement settings~\cite{vonNeumann1955,4h4b-3xss,doi:10.1142/S0219477525400280}. Among these, quantum metrology exploits nonclassical resources to achieve measurement sensitivities beyond those attainable with classical devices ~\cite{giovannetti2011advances,degen2017quantum}. A key resource enabling such advantages is quantum entanglement, which can enhance the performance of sensing and detection protocols ~\cite{horodecki2009quantum,maleki2021quantum}.

Within this framework, quantum illumination (QI) has emerged as a paradigmatic quantum sensing protocol in which entangled states are used to enhance the detection of weakly reflecting targets embedded in noisy environments~\cite{shapiro2009quantum,sacchi2005entanglement}. QI has found applications in quantum communication~\cite{shapiro2009defeating}, quantum radar~\cite{karsa2024quantum,blakey2022quantum,liu2023compact}, and quantum cloaking~\cite{las2017quantum}. In QI, the probe consists of two correlated modes, commonly referred to as the signal and idler: the signal is transmitted toward the sensing region to interrogate the presence of a target, while the idler is retained locally. After the interaction with the environment, the returning signal is jointly measured with the stored idler at the receiver. The task is to discriminate between the target-present and target-absent hypotheses by minimizing the error probability based on the measurement outcomes~\cite{helstrom1969quantum,barnett2009quantum,bae2015quantum,calsamiglia2008quantum,pirandola2008computable,PhysRevLett.118.070803}, which is bounded by the Helstrom limit~\cite{sacchi2005optimal,lloyd2008enhanced} and the quantum Chernoff bound (QCB)~\cite{audenaert2007discriminating}.

QI protocols have been extensively developed using Gaussian probe states, most notably the two-mode squeezed state (TMSS)~\cite{tan2008quantum}, an entangled Gaussian resource that has been widely studied both theoretically and experimentally~\cite{guha2009gaussian,zhang2015entanglement,zhuang2017optimum}. The TMSS has been shown to outperform optimal classical illumination (CI) based on coherent states under the same energy constraint, particularly in the presence of bright environmental noise. All these studies indicate that the robustness of QI against environmental noise is intrinsically linked to the entanglement between the signal and idler, whose effectiveness is determined by how it is encoded into measurable inter-mode correlations at the detection stage~\cite{shapiro2020quantum}. While Gaussian states such as the TMSS provide a significant advantage over CI, they are not necessarily optimal for exploiting quantum advantages in QI protocols. Consequently, considerable effort has been devoted to engineering probe states with tailored correlation structures. In this context, local non-Gaussian protocols have attracted particular attention, with operations such as photon addition (PA), photon subtraction (PS), and photon catalysis (PC) providing effective routes to enhance correlations and improve detection performance in both noiseless and noisy environments~\cite{ourjoumtsev2006increasing,yang2009entanglement,kim2013enhanced}.

Several studies have investigated the impact of local non-Gaussian operations on the performance of quantum illumination. Early work on PS demonstrated that it ca reduce the error probability in the discrimination task~\cite{zhang2014ps} and remain robust under photon-loss channels~\cite{zhang2014quantum}. Subsequent investigations considered more general state-engineering strategies, including combinations of photon subtraction and photon addition, showing improved performance over the baseline TMSS transmitter~\cite{fan2018quantum}. More recently, alternative approaches such as zero-photon subtraction (ZPS) have been proposed, exhibiting enhanced performance in quantum radar scenarios even under realistic conditions of loss~\cite{zhang2025enhancing}. In parallel, systematic comparisons of local non-Gaussian protocols, including PS, PA, and PC, have confirmed their advantages over CI and further shown that appropriate joint measurement strategies, such as balanced homodyne detection, can significantly enhance the signal-to-noise ratio (SNR) in target detection~\cite{zhang2024quantum}.

In various studies, PC has been identified as an effective local non-Gaussian operation capable of enhancing both entanglement and success probability relative to photon addition and photon subtraction. However, when applied to the TMSS, its success probability remains limited, typically below $20\%$, while the entanglement enhancement remains relatively significant. To overcome this limitation, Ref.~\cite{liu2022optimal} proposed a nonlocal non-Gaussian photon addition (NLPA) incorporating an additional beam splitter together with PA, achieving success probabilities exceeding $70\%$ while maintaining strong entanglement enhancement.

In this paper, we consider the use of the NLPA protocol as an engineered probe state within the QI protocol. Our results show that this engineered state outperforms conventional PC, PA, and PS transmitters while requiring only a single auxiliary photon. Furthermore, it exhibits robust entanglement over a broad range of beam-splitter parameter, reducing the need for precise parameter tuning, and can be effectively implemented thanks to its high success probability, providing a practical advantage for discrimination tasks even in the presence of photon-loss channels. 

The paper is organized as follows. Section~\ref{sec2} introduces the QI discrimination framework. Section~\ref{sec3} presents the QI protocol under non-Gaussian operations, including both local (Sec.~\ref{sec3a}) and nonlocal (Sec.~\ref{sec3b}) schemes. We then analyze the performance with one and two auxiliary photons (Secs.~\ref{sec3c} and~\ref{sec3d}), as well as its robustness under photon loss and the receiver design (Secs.~\ref{sec3e} and~\ref{sec3f}). Conclusions and future directions are given in Sec.~\ref{sec4}.

\section{Quantum Illumination: General Framework} 
\label{sec2}

The central idea of QI is to exploit quantum correlations to detect weakly reflecting targets in noisy environments immersed in a thermal background. Considering quantum advantage, a quantum protocol requires the preparation of a quantum resource state, such as an entangled state composed of two modes, denoted $A$ and $B$. Mode $A$ (idler) is retained locally at the receiver, while mode $B$ (signal), is transmitted toward the region where the potential target may be located. The interaction of the signal mode with the environment and the possible target encodes the information required to infer the presence of the object.


The interaction with the target can be modeled as a beam splitter with weak reflectivity $\kappa \ll 1$. In this description, the signal mode $B$ mixes with an environmental thermal mode $C$ representing the background noise, characterized by an average photon number $\bar{n}_{\mathrm{th}}$. After this interaction, the reflected mode $B'$ returns to the receiver, where it is jointly measured together with the retained idler mode $A$ in order to infer the presence or absence of the target.

During this interaction, when the target is present, the returning mode results from the mixing of the signal mode $B$ with the environmental thermal mode $C$ through the beam-splitter unitary transformation $\hat U_{\mathrm{BS}} = e^{\theta G}$, where $\theta = \arctan\!\left(\sqrt{\frac{1-\kappa}{\kappa}}\right)$ and $G = \hat a_{B}\hat a_{C}^{\dagger} - \hat a_{B}^{\dagger}\hat a_{C}$. Here, $\hat a_{B}$ and $\hat a_{C}$ denote the annihilation operators associated with the signal and thermal noise modes, respectively. In this case, the average photon number of the thermal environment is taken to be $\bar{n}'_{\mathrm{th}} = \frac{\bar{n}_{\mathrm{th}}}{1-\kappa}$ so that the returning mode exhibits an effective background photon number $\bar{n}_{\mathrm{th}}$ \cite{tan2008quantum,shapiro2009quantum}. However, when the target is absent, the transmitted signal is completely lost, and the returning mode consists only of the environmental thermal noise. Consequently, the detection problem can be formulated as a binary hypothesis test characterized by two possible output states, $\rho_0$ and $\rho_1$ which is inferred at the measurement stage. 

Under the null hypothesis $H_0$ (target absent), the joint state of the retained idler and the returned mode is given by
\begin{equation}
\rho_{0}
= \rho_{A} \otimes \rho_{C}(\bar{n}_{\mathrm{th}}),
\end{equation}
where $\rho_{C}(\bar{n}_{\mathrm{th}})$ denotes a thermal state with an average photon number $\bar{n}_{\mathrm{th}}$. Here, $\rho_{A}$ is the idler state that is obtained from the reduced density operator of the initial bipartite state $\rho_{AB}$,
\(\rho_{A} = \Tr_{B}(\rho_{AB})
\).

Under the alternative hypothesis $H_1$ (target present), the signal interacts with the environment through the beam-splitter transformation described previously. After tracing over the environmental mode $C$, the joint state of the idler and the returning mode becomes \cite{tan2008quantum}
\begin{equation}
\rho_{1}
= \Tr_{C} \!\left[
\hat U_{\mathrm{BS}}(\kappa)
\big(
\rho_{AB} \otimes \rho_{C}(\bar{n}'_{\mathrm{th}})
\big)
\hat U_{\mathrm{BS}}^{\dagger}(\kappa)
\right]. \label{rho1}
\end{equation}

After reception, the returned signal is jointly measured with the stored idler to discriminate between the two possible states $\rho_0$ and $\rho_1$, corresponding to target absence and presence. The performance is quantified by the error probability under equal a priori probabilities, reducing the problem to minimum-error state discrimination.

When $K$ identical copies of the signal-idler pairs are available, the optimal error probability is given by the Helstrom limit~\cite{helstrom1969quantum,sacchi2005optimal,lloyd2008enhanced},
\begin{equation}
P_{\mathrm{error},K}
=\frac{1}{2}\!\left(1-\frac{1}{2}\left\|\rho_1^{\otimes K}-\rho_0^{\otimes K}\right\|_1\right). \label{Helmostlimit}
\end{equation}
In realistic scenarios, evaluating the Helstrom limit becomes computationally demanding, since the dimension of the joint states $\rho_0^{\otimes K}$ and $\rho_1^{\otimes K}$ grows exponentially with the number of copies $K$. Moreover, the trace norm appearing in the Helstrom expression does not generally admit a simple analytic evaluation for tensor powers of mixed states. For this reason, it is common to employ the quantum Chernoff bound (QCB), which provides an asymptotically tight and computationally convenient upper bound on the error probability in the large-$K$ limit~\cite{audenaert2007discriminating,tan2008quantum}. The QCB bound is given by
\begin{equation}
P_{\mathrm{error},K}\leq \frac{1}{2}Q^K , \label{QCB}
\end{equation}
where 
\begin{equation}
  Q=\min_{0\leq s\leq 1}\Tr\!\left(\rho_0^{\,s}\rho_1^{\,1-s}\right).  \label{Q}
\end{equation}


\section{Quantum Illumination with non-Gaussian Entangled States}
\label{sec3}

\subsection{Local non-Gaussian Entangled States}
\label{sec3a}

Within the framework of continuous-variable quantum information processing \cite{RevModPhys.77.513,RevModPhys.84.621}, TMSS is one of the most natural quantum resource states and has found broad application in a variety of settings. The TMSS is defined as \cite{scully1997quantum}

\begin{equation}
\ket{\psi}_{AB}
=\sqrt{1-\lambda^2}\sum_{n=0}^{\infty}\lambda^n\,\ket{n}_{A}\ket{n}_{B},\label{TMSS}
\end{equation}
where $\lambda=\tanh r$ and $r$ is the squeezing parameter. This state exhibits quantum correlations between the signal and idler modes and has been widely studied as a benchmark resource for quantum illumination \cite{guha2009gaussian,zhang2015entanglement,zhuang2017optimum}. 

A key ingredient underlying the performance of QI is the presence of entanglement and nonclassical correlations between the transmitted signal and the retained idler. This has motivated the engineering of probe states with enhanced or modified entanglement properties, particularly through local non-Gaussian states and entanglement-enhancement protocols, as promising resources for improved detection performance~\cite{fan2018quantum,zhang2024quantum}. Among the most widely studied local non-Gaussian operations are photon subtraction (PS), photon addition (PA), and photon catalysis (PC). Various combinations of photon-subtraction and photon-addition operations have been shown to enhance the entanglement of Gaussian states \cite{ourjoumtsev2006increasing,yang2009entanglement}.

\begin{figure}[!htbp]
\includegraphics[width=0.95\linewidth]{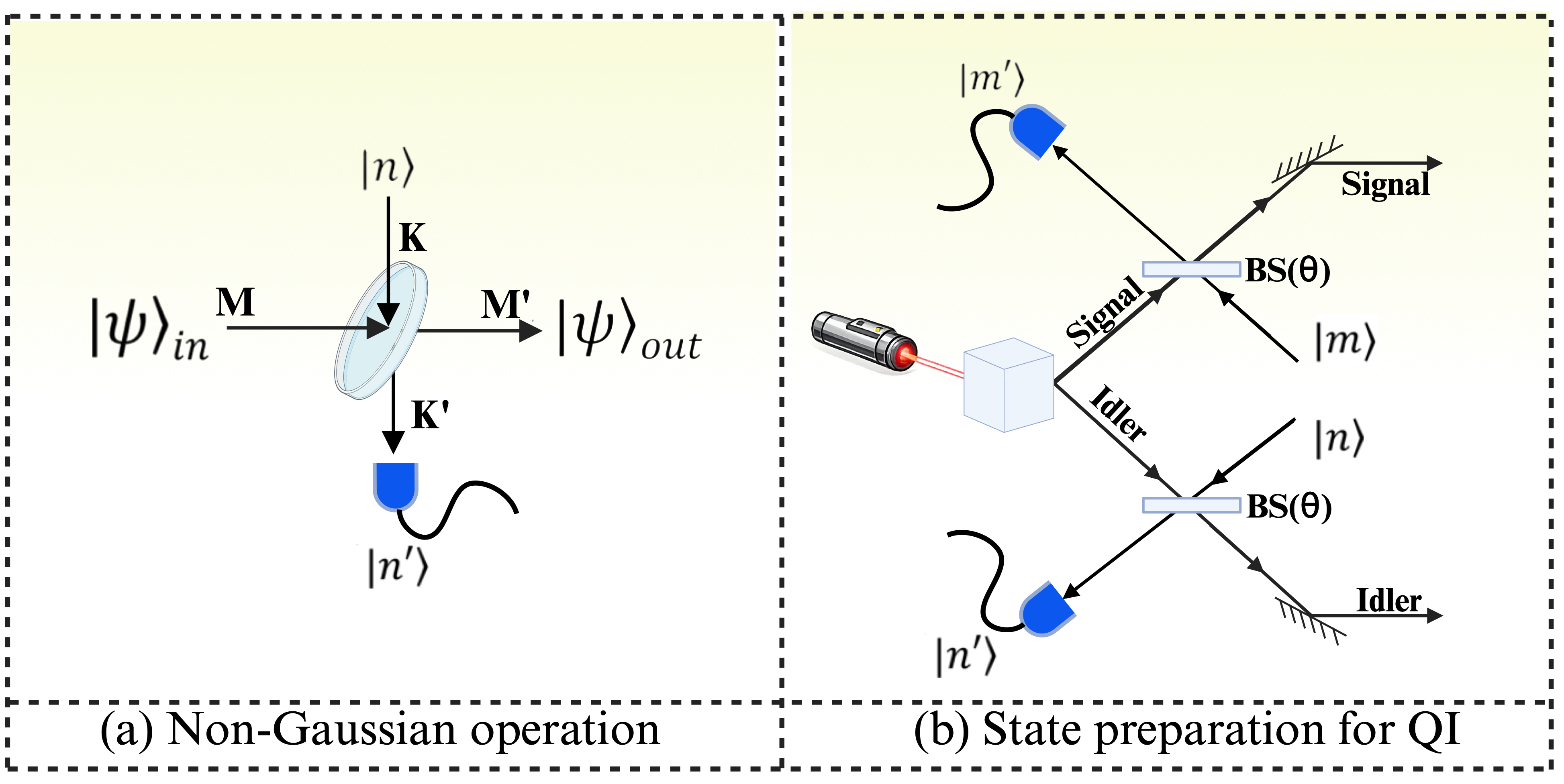}
\caption{
(a) Conditional state preparation via beam-splitter interaction. 
(b) State preparation using local non-Gaussian operations with TMSS input.
}
\label{sch3}
\end{figure}

To illustrate the implementation of local non-Gaussian operations, we consider a general conditional protocol shown in Fig.~\ref{sch3}(a), where an input mode $M$ interferes at a beam splitter of transmissivity $T=\cos^2\theta$ with an auxiliary Fock state $\ket{n}$. Conditioning on the detection of $n'$ photons at one output port realizes different operations: photon addition for $n>n'$, photon subtraction for $n<n'$, and photon catalysis for $n=n'$. In this setting, the beam splitter unitary transformation can be written as
\begin{equation}
\hat{U}_{\mathrm{BS}}(T)=\exp\!\left[\theta\left(\hat a_{K}^{\dagger}\hat a_{M}-\hat a_{K}\hat a_{M}^{\dagger}\right)\right],
\end{equation}
where $\hat a_{M}$ and $\hat a_{K}$ denote the annihilation operators associated with the input mode and the auxiliary mode, respectively. Within this conditional-measurement protocol, the action of the beam splitter followed by photon-number detection defines an effective operation on the input state. For an input state of the form $\ket{\psi}_{\mathrm{in}}=\sum_k c_k \ket{k}$, the resulting state can be written as
\begin{equation}
\ket{\psi}_{\mathrm{out}}=\hat B_{n,n'}\,\ket{\psi}_{\mathrm{in}},
\end{equation}
where the operator is defined as
\begin{equation}
\hat B_{n,n'}=\bra{n'}\hat{U}_{\mathrm{BS}}(T) \ket{n}=\sum_{k=0}^{\infty} B_{n,n',k}\ket{k+n-n'}\bra{k}.
\end{equation}
Consequently, the normalized output state becomes
\begin{equation}
    \ket{\psi}_{\mathrm{out}}=N_{n,n'}\sum_{k=0}^{\infty}c_k B_{n,n',k} \ket{k+n-n'}, \label{ngoequ}
\end{equation}
where $N_{n,n'}$ is a normalization constant. The coefficients $B_{n,n',k}$ are given by~\cite{liu2022optimal}
\begin{equation}
\begin{aligned}
B_{n,n',k}
= \sum_{i=0}^{n} (-1)^{\,n'-i}
\binom{n}{i}
\binom{k}{n'-i}
(\sqrt{T})^{\,k+2i-n'} \\
\times (\sqrt{1-T^2})^{\,n+n'-2i}
\frac{\sqrt{(k+n-n')!\, n'!}}{\sqrt{k!\,n!}} .
\end{aligned} \label{Bcoef}
\end{equation}
The pair $(n,n')$ specifies the type of local non-Gaussian operation realized by the conditional measurement. For instance, $(1,0)$ corresponds to a single photon addition, $(0,1)$ corresponds to  a single photon subtraction, and $(1,1)$ corresponds to photon catalysis. Various studies have applied the above local non-Gaussian operation to enhance the entanglement of quantum states in QI, such as squeezed states, and thereby improve quantum performance, either by acting on a single mode or jointly on both modes, as illustrated in Fig.~\ref{sch3}(b)~\cite{ourjoumtsev2006increasing,yang2009entanglement,kim2013enhanced}.

\subsection{Nonlocal Non-Gaussian Photon Addition}
\label{sec3b}

\begin{figure*}[!htbp]
\centering
\includegraphics[width=0.8\textwidth]{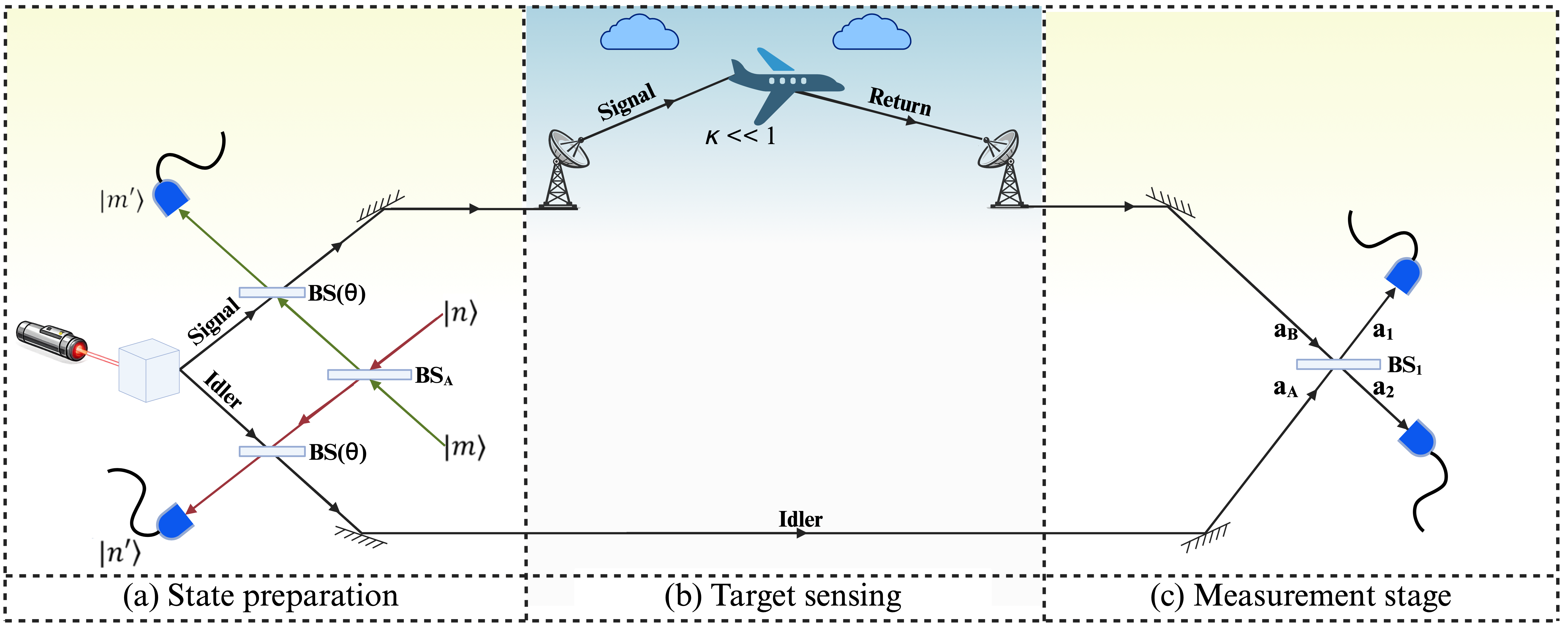}
\caption{
Schematic representation of the QI protocol using NLPA. 
(a) State preparation: a TMSS source generates signal ($B$) and idler ($A$) modes that interact with auxiliary Fock states $|m\rangle$ and $|n\rangle$ through beam splitters $\mathrm{BS}_A$ and $\mathrm{BS}(\theta)$, followed by conditional detection of $\ket{m'}$ and $\ket{n'}$, producing the non-Gaussian transmitter. 
(b) Target sensing: the signal mode propagates toward a weakly reflecting target characterized by transmissivity $\kappa \ll 1$ in a noisy environment, while the idler mode is retained locally. (c) Measurement stage: The returned mode and the stored idler are interfered on a $50{:}50$ beam splitter $\mathrm{BS}_1$. Photon-number measurements are performed at both outputs, and their difference $\hat M_{\mathrm{BS}} = \hat n_1 - \hat n_2$ is used as the decision statistic.}
\label{sch1}
\end{figure*}

An effective strategy for entanglement enhancement is obtained by modifying the standard conditional local non-Gaussian protocol through an additional beam splitter acting on the auxiliary modes prior to their interaction with the TMSS ~\cite{liu2022optimal}. As shown in Ref.~\cite{liu2022optimal}, this configuration, see Fig.~\ref{sch1}(a), outperforms photon subtraction, addition, and catalysis applied directly to the TMSS, both in entanglement and success probability. 

To consider this version of coherent photon addition, we start from the TMSS state in Eq.~\eqref{TMSS}. The protocol includes auxiliary number states with $m=1$ and $n=0$ in Fig.~\ref{sch1}(a), corresponding to nonlocal non-Gaussian photon addition, followed by vacuum detection in the auxiliary output ports. This realizes an effective single-photon addition process that acts coherently on the two-mode resource. 


In this protocol, the resulting output state can be written as
\begin{equation}
\ket{\psi_1}
=
\sum_{n=0}^{\infty} \sqrt{p_n}\,\ket{\phi_n}, \label{psif}
\end{equation}
with
\begin{equation}
\ket{\phi_n}
=
\frac{1}{\sqrt{2}}
\left(
\ket{n+1}_{A}\ket{n}_{B}
+
\ket{n}_{A}\ket{n+1}_{B}
\right), \label{phin}
\end{equation}
where $p_n
=
\left(1-\lambda^2T^2\right)^2
(\lambda T)^{2n}(n+1)$ and $T=\cos^2 \theta$. This representation is particularly revealing because each basis element $\ket{\phi_n}$ is a maximally entangled state within the corresponding two-dimensional excitation subspace. Therefore, the output state is a coherent superposition of maximally entangled components whose amplitudes are controlled by the effective parameter $\lambda T$.  
The coefficients $\{p_n\}$ define the probability weight associated with each maximally entangled component.

\subsection{Single-Auxiliary-Photon Case} \label{sec3c}

Now, we analyze the performance of QI using the NLPA protocol, shown in Fig.~\ref{sch1}, with a single auxiliary photon ($m=1$, $n=0$), and compare it with local non-Gaussian operations (PA, PC, and PS) over the signal mode under an equivalent resource constraint which has been considered in the previous studies \cite{fan2018quantum,zhang2025enhancing,zhang2024quantum}. Specifically, a single auxiliary photon is injected in the signal mode ($m=1$) for PA and PC, while one photon is removed in the case of PS ($m'=1$). Unless otherwise stated, all results are obtained with $\bar{n}_{\mathrm{th}}=1$, $\kappa=0.01$, and $\sinh^2(r)=0.05$. For numerical evaluation, the photon-number space is truncated at $n_{\max}=24$ for all considered states. As a first step, we quantify the entanglement generated by these schemes, motivated by its well-established connection to QI performance and its role as a fundamental quantum resource. Specifically, we evaluate the von Neumann entropy of the reduced density operator associated with the signal. The reduced state is obtained by tracing out the idler subsystem, $\rho_{B}=\Tr_{A}(\rho_{AB})$.
The entanglement entropy is then given by \cite{bennett1996concentrating}
\begin{equation}
E_V(\rho_{\mathrm{B}})
=
-\Tr(\rho_B\,\log_2\,\rho_B). \label{Ev}
\end{equation}
In addition to entanglement, it is essential to evaluate the success probability associated with the NLPA protocol and the corresponding local non-Gaussian operations. We first find the success probability for the NLPA as ~\cite{liu2022optimal}
\begin{equation}
    P_{\mathrm{NLPA}}=\frac{\operatorname{sech}^2 r\,(1-T)}{(1-\lambda^2 T^2)^2}. \label{PNLPA}
\end{equation}

Similarly, the success probability for a local non-Gaussian operation over one mode is given by
\begin{equation}
    P_{\mathrm{nGo}}=\sum_{k=0}^{\infty}|C_k|^2, \label{PnGo}
\end{equation}
where the coefficients $C_k$ take the form
\begin{equation}
    C_k=\operatorname{sech} r \,\lambda^k \,B_{m,m',k}.
\end{equation}

The coefficients $B_{m,m',k}$ are defined by Eq.~\eqref{Bcoef}. These expressions fully characterize the photon-number weights generated by the local non-Gaussian transformations and the entanglement-enhancing scheme and therefore determine the corresponding success probabilities. 

\begin{figure}[!htbp]
\centering
\includegraphics[width=\linewidth]{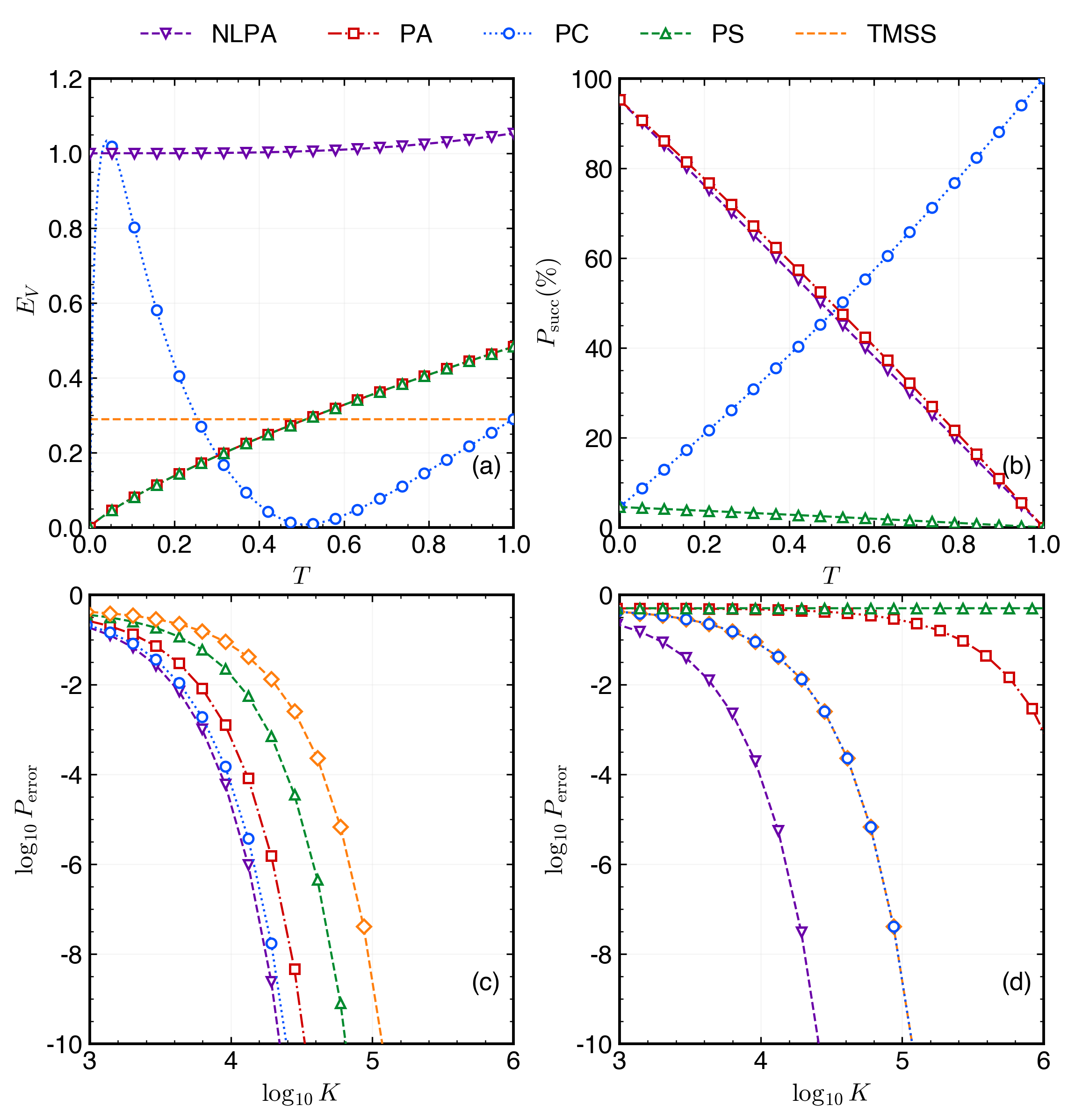}
\caption{
(a) Von Neumann entropy $E_V$ of the reduced state $\rho_B$  and (b) Success probability $P_{\mathrm{succ}}$ as a function of the beam splitter transmissivity $T$. Error probability as fuction of number of copies $K$ (c) evaluated at the transmissivity $T$ that maximizes the von Neumann entropy $E_V$ and (d) transmissivity $T$ that maximizes the success probability for each protocol using one auxiliary photon. The TMSS case is shown as a reference.}
\label{fig1}
\end{figure}

In Fig.~\ref{fig1}(a) we present the von Neumann entropy as a function of the transmissivity $T$ of the beam splitter. From Fig.~\ref{fig1}(a), it is observed that the NLPA scheme yields the largest entropy, reaching a maximum value of approximately $1.06$. A notable feature is its stability, maintaining values close to unity over a broad range of $T$. In contrast, PC attains a slightly lower peak value of about $1.03$, but rapidly decreases as $T$ increases, eventually falling below the entanglement of the TMSS in the limit $T \to 1$, thereby restricting the parameter regime over which enhancement is achieved. On the other hand, PA and PS initially produce entanglement below that of the TMSS (approximately $0.28$), but they increase monotonically with $T$, reaching a maximum value of about $0.483$ as $T \to 1$~\cite{zhang2024quantum}. This behavior indicates only a limited enhancement relative to the TMSS for these operations.

In a realistic implementation of engineered operations for QI protocols, the success of the protocol cannot be analyzed independently of the associated success probability, as the engineered-state preparation is inherently probabilistic. Its practical relevance, therefore, depends critically on the associated success probability. In Fig.~\ref{fig1}(b), we compare the success probabilities of the four schemes. The NLPA protocol exhibits a high initial success probability which gradually decreases and becomes negligible as $T \to 1$. A key feature of this behavior is that, over a broad range where the success probability remains above $70\%$, the corresponding entanglement [as shown in the Fig~\ref{fig1}(a)] is also significant ~\cite{liu2022optimal}. In contrast, PA, PS, and PC do not exhibit a favorable correspondence between high entanglement and high success probability. In particular, PC remains below $20\%$ in the region where it achieves its maximum von Neumann entropy ~\cite{liu2022optimal}. Similarly, PA and PS reach their highest entanglement values only in regimes where the success probability is vanishingly small. This imposes a significant limitation on local non-Gaussian operations, as the most probable outcome is that they do not provide a considerable improvement over the initial state in the QI protocol.

To elucidate the interplay between entanglement (quantified by the von Neumann entropy), success probability, and QI performance (measured by the error probability), we present the results in Figs.~\ref{fig1}(c) and (d). Panel (c) shows the error probability as a function of the number of signal-idler copies $K$, evaluated at the transmissivity $T$ that maximizes the von Neumann entropy-namely, $T \to 1$ for PA, NLPA, and PS, and $T = 0.041$ for PC. Panel (d) displays the error probability as a function of $K$ at the transmissivity that maximizes the success probability, corresponding to $T \to 0$ for PA, NLPA, and PS, and $T \to 1$ for PC.

From Fig.~\ref{fig1}(c), it is evident that NLPA protocol offers the lowest values of error probability. Since reduced error probabilities correspond to improved target-detection performance in QI, this indicates that NLPA provides the best performance among the considered schemes. PC follows closely, exhibiting a comparable error probability with only a marginal difference relative to NLPA. Although PA and PS also achieve lower error probabilities than TMSS, their advantage is significantly limited by the associated success probabilities. These analysis agree with previous studies reported in \cite{fan2018quantum,zhang2025enhancing,zhang2024quantum}. In all cases, the maximum von Neumann entropy, at which these curves are evaluated, occurs in a regime where the success probability is very low, typically not exceeding $10\%$. As a result, the apparent performance enhancement corresponds to a highly unlikely event and therefore are not favorable for the practical improvements.

We consider in Fig.~\ref{fig1}(d) the error probability as a function of $K$, evaluated at values of $T$ yielding high success probability. The results show that NLPA again achieves the lowest error probability, and is the only scheme that outperforms TMSS in this regime.  In contrast, the local non-Gaussian operations do not provide a meaningful advantage: PC approaches the TMSS performance, while PA and PS exhibit no improvement and, in fact, lead to higher error probabilities. These results indicate that, under likely operating conditions, a genuine quantum advantage is obtained only with NLPA.

Next, we consider the error exponent~\cite{zhang2025enhancing,assouly2023quantum}, which provides an asymptotic measure of the QI performance by quantifying the exponential rate at which the error probability decays as the number of repeated measurements increases. The error exponent is defined as ~\cite{zhang2025enhancing,assouly2023quantum}
\begin{equation}
    \epsilon=\lim_{K\to \infty} -\frac{1}{K}\ln P_{\mathrm{error}},
\end{equation}
which implies that $P_{\mathrm{error}}$ is logarithmically equivalent to $e^{-\epsilon K}$. A larger value of $\epsilon$ indicates a faster decay of the error probability and therefore a more efficient discrimination between the two hypotheses. Using Eqs.~\eqref{QCB}, the corresponding error exponents are given by ~\cite{zhang2025enhancing}
\begin{subequations}
\begin{align}
    \epsilon_{\alpha} &= -P_{\alpha} \ln Q_{\alpha},\\
    \epsilon_{\mathrm{TMSS}}  &= - \ln Q_{\mathrm{TMSS}}\label{tmep},
\end{align}
\end{subequations}
with $\alpha \in \{\mathrm{NLPA}, \mathrm{PA}, \mathrm{PC}, \mathrm{PS}\}$ labeling the applied non-Gaussian operation and $Q_{\alpha}$ is defined in Eq.~\eqref{Q}, and $P_{\alpha}$ is defined in Eqs.~\eqref{PNLPA} and \eqref{PnGo}. To quantify the advantage of NLPA over local non-Gaussian operations and TMSS, we consider the gain ratio
\begin{equation}
G_{\alpha}=\frac{\epsilon_{\alpha}}{\epsilon_{\mathrm{TMSS}}}, \label{Gratio}
\end{equation}
where $G_{\alpha}>1$ indicates an improvement over the TMSS when the probabilistic nature of the state engineering is taken into account.

\begin{figure}[!htbp]
\centering
\includegraphics[width=\columnwidth]{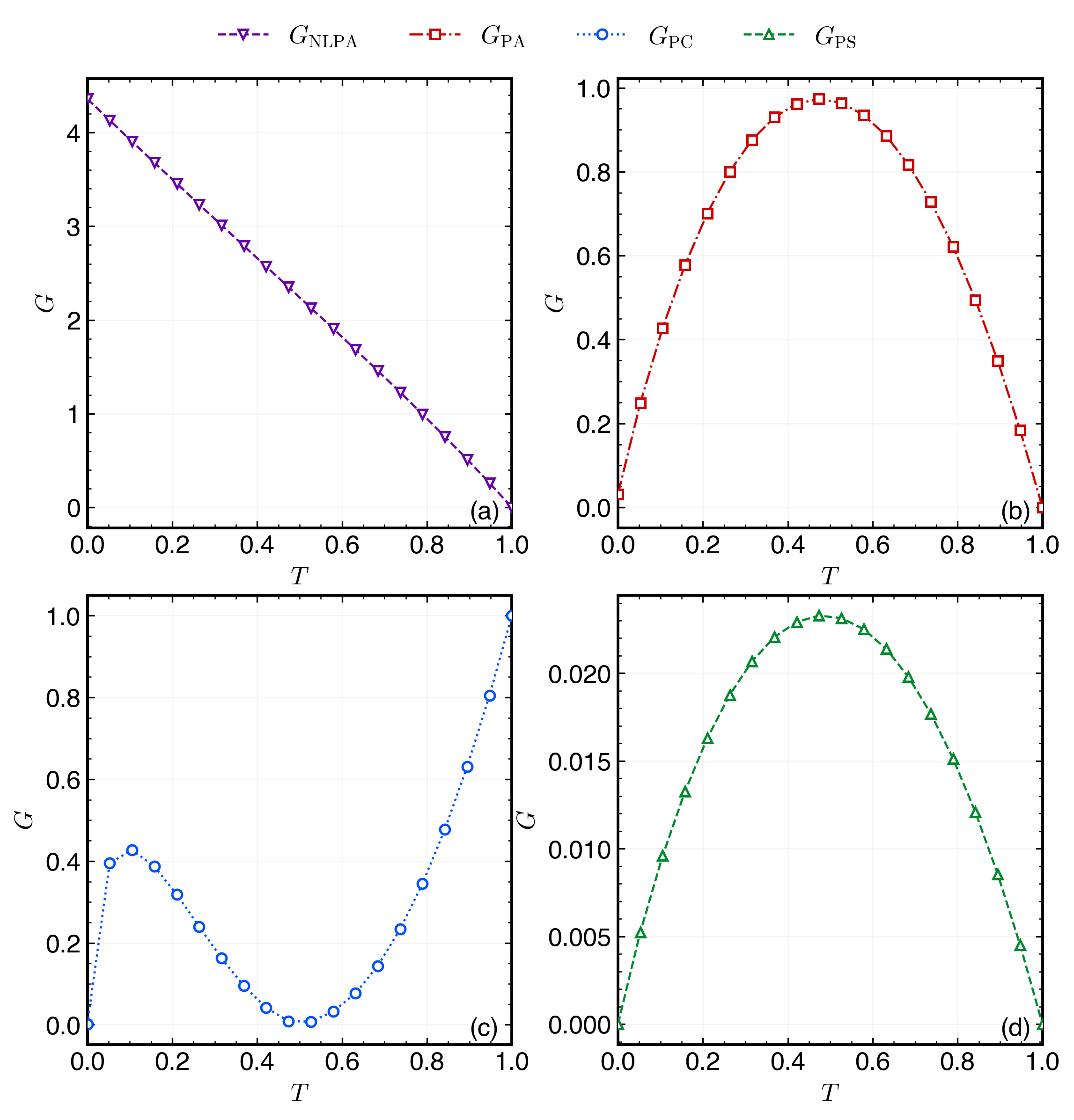}
\caption{
Error exponent ratio $G$ as a function of transmissivity $T$ using one auxiliary photon. Panels correspond to the ratio of (a) nonlocal photon addition (NLPA), (b) local photon addition (PA), (c) local photon catalysis (PC), and (d) local photon subtraction (PS). }\label{fig3}
\end{figure}

In Fig.~\ref{fig3}, we present the gain parameter $G$ as a function of the beam-splitter transmissivity $T$ for the four protocols considered. The behavior is strongly protocol dependent and reveals that the effective advantage encoded in $G$ does not follow a universal monotonic trend with $T$. Figure~\ref{fig3}(a) shows that $G_{\mathrm{NLPA}}$ decreases monotonically with $T$, taking its largest values in the low-$T$ region and vanishing as $T\rightarrow1$. This indicates that the effective exponent associated with NLPA is maximized when the heralding probability remains large, whereas the probabilistic cost of state preparation progressively suppresses the gain as $T$ increases. In Fig.~\ref{fig3}(b), $G_{\mathrm{PA}}$ exhibits a broad concave profile, reaching a maximum near intermediate transmissivities, $T=0.5$, and remaining below unity throughout the full range. This shows that, once the success probability is incorporated into the effective exponent, photon addition does not attain an advantage comparable to the strongest NLPA setting. The intermediate-$T$ maximum reflects a compromise between improved channel distinguishability and the decreasing probability of successful conditional preparation. The behavior of $G_{\mathrm{PC}}$ in Fig.~\ref{fig3}(c) is markedly non-monotonic. It is moderate at small $T$, decreases to a shallow minimum around intermediate transmissivities, and then increases again as $T\to1$, approaching unity in the high-$T$ limit. This indicates that the catalysis protocol is highly sensitive to the balance between state improvement and success probability. In particular, $G_{\mathrm{PC}}$ near $T\to1$ suggests that, in this limit, the effective exponent becomes comparable to the chosen reference as we see in~Fig.\ref{fig1}(d). Finally, Fig.~\ref{fig3}(d) shows that $G_{\mathrm{PS}}$ remains very small over the entire transmissivity range, with only a shallow maximum around $T=0.5$ and vanishing values at both $T=0$ and $T=1$. This demonstrates that photon subtraction yields the weakest effective gain among the four cases considered. Even when subtraction enhances nonclassical correlations at the state level, the corresponding heralding penalty strongly suppresses its operational advantage.

Overall, the figure shows that the parameter $G_{\alpha}$ is controlled by a nontrivial competition between discrimination capability and success probability. Since the effective exponent contains the product $-P_{\alpha}\ln Q_{\alpha}$, a protocol may exhibit favorable distinguishability properties while still yielding a small gain if the conditional preparation probability is strongly reduced. For this reason, the trends in $G_{\alpha}$ should be interpreted as operational, success-probability-weighted performance rather than as a direct measure of conditional-state quality alone.

\subsection{Two-Auxiliary-Photon Case} \label{sec3d}

In this subsection, we extend the analysis of the non-Gaussian protocols to the scenario of two auxiliary photons. In this case, the NLPA scheme corresponds to the configuration $(m=n=1)$, 
with the engineered state
\begin{equation}
    \ket{\psi_2}=\sum_{n=0}^{\infty} \sqrt{q_n} \,\ket{\chi_n},
\end{equation}
with
\begin{equation}
    q_n=\frac{(1-\lambda^2T^2)^3}{2}(\lambda T)^{2n}(n+1)(n+2),
\end{equation}
and
\begin{equation}
   \ket{\chi_n}=\frac{1}{\sqrt{2}}\left(\ket{n}_{\mathrm{A}}\ket{n+2}_{\mathrm{B}}-\ket{n+2}_{\mathrm{A}}\ket{n}_{\mathrm{B}}\right).
\end{equation}

The associated success probability is given by
\begin{equation}
P_{\mathrm{NLPA}}=\frac{\operatorname{sech}^2 r\,(1-T)^2}{(1-\lambda^2 T^2)^3}. \label{PNLPA2}
\end{equation}

Similarly, for the local non-Gaussian operations applied to both modes with $(m=n=1)$, the success probability, assuming equal transmissivity, is given by $P_{\mathrm{nGo}}=\sum_{k=0}^{\infty}|C_k|^2, $
where $   C_k=\operatorname{sech} r \,\lambda^k \,B_{m,m',k}^2.$

The coefficients $B_{m,m',k}$ are defined in Eq.~\eqref{Bcoef}. The entanglement is quantified again using the von Neumann entropy defined in Eq.~\eqref{Ev}. 

\begin{figure}[!htbp]
\centering
\includegraphics[width=\linewidth]{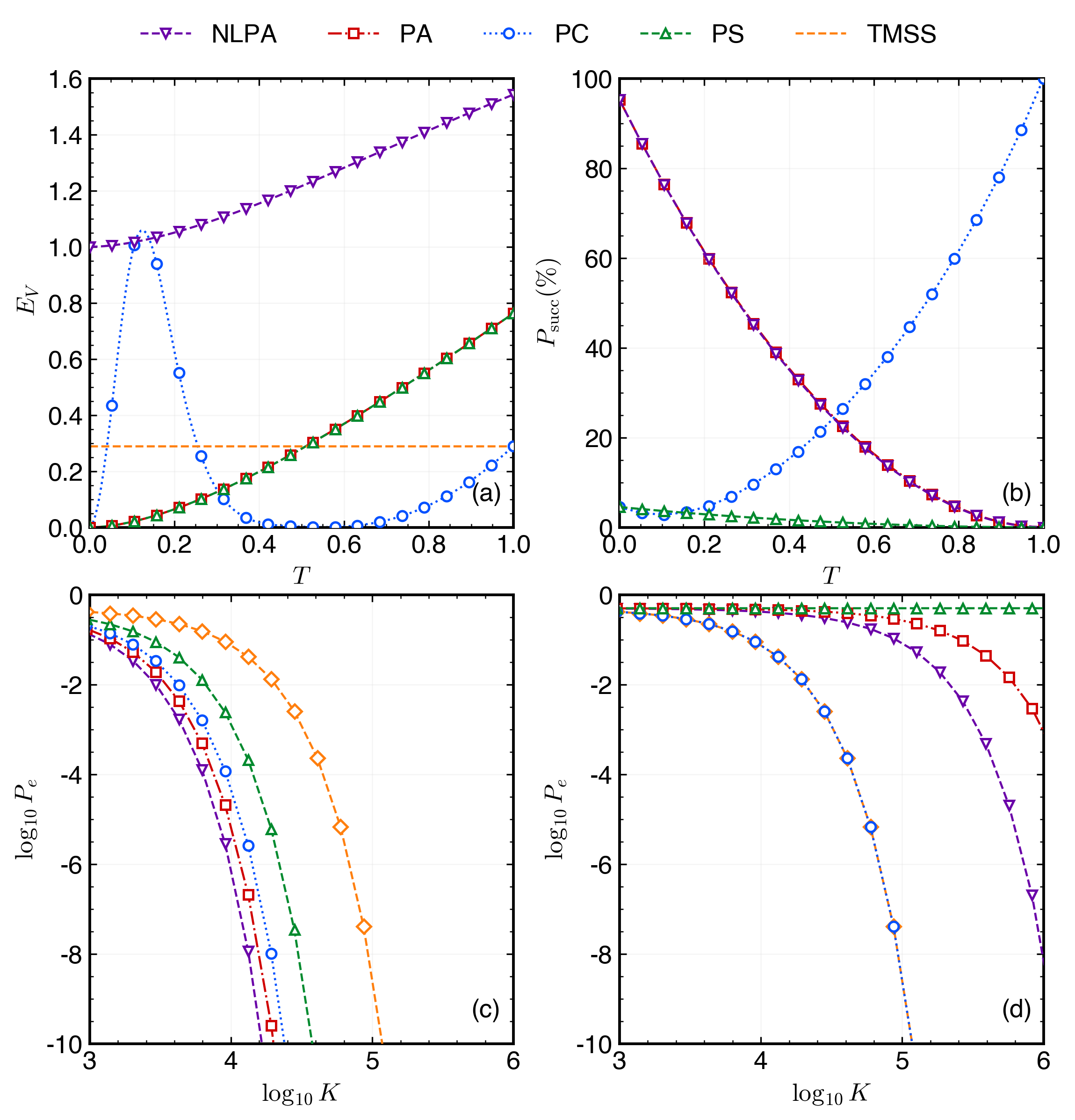}
\caption{
(a) Von Neumann entropy $E_V$ of the reduced state $\rho_B$  and (b) Success probability $P_{\mathrm{succ}}$ as a function of the beam splitter transmissivity $T$. Error probability as fuction of number of copies $K$ (c) evaluated at the transmissivity $T$ that maximizes the von Neumann entropy $E_V$ and (d) transmissivity $T$ that maximizes the success probability for each protocol using two auxiliary photons. The TMSS case is shown as a reference.
}
\label{fig4}
\end{figure}

In Figs.~\ref{fig4}(a) and (b), we present the von Neumann entropy and the corresponding success probability, as functions of the transmissivity $T$, respectively. From Fig.~\ref{fig4}(a), the NLPA protocol offers the highest entanglement, reaching values close to $1.5$ in the limit $T \to 1$. As in the single-auxiliary-photon case, NLPA remains the most stable scheme, maintaining a higher entropy over the entire range of $T$. This indicates that the engineered state is dominated by highly entangled components across all regimes. In contrast, PC achieves a lower maximum value of approximately $1.06$ compared with NLPA, followed by a rapid decrease that eventually falls below the TMSS. The PA and PS protocols exhibit similar behavior: they start from lower values and increase monotonically with $T$, reaching a maximum of approximately $0.763$ as $T \to 1$~\cite{zhang2024quantum}.

As discussed previously, entanglement alone is not sufficient to assess the practical performance of all protocols and must be considered together with the success probability. The corresponding results are shown in Fig.~\ref{fig4}(b). The behavior of the NLPA protocol remains similar to that observed in the single-photon case [See Fig.~\ref{fig1}(b)], with an initial success probability of approximately $95\%$, followed by a more rapid decrease as $T \to 1$, becoming negligible for $T \gtrsim 0.8$. In contrast, PC exhibits an opposite trend, increasing from near zero to its maximum value as $T \to 1$. PA follows a behavior similar to the NLPA scheme, consistent with the fact that both involve photon addition processes, although the NLPA consistently achieves higher entanglement. Finally, PS maintains low success probabilities, not exceeding $10\%$ over the entire range. The key observation is that NLPA sustains high entanglement in a regime where the success probability remains above $60\%$, making it both effective and experimentally relevant. In contrast, for the remaining protocols, their maximum entanglement occurs only in regimes of low success probability, corresponding to unlikely events. As a result, these local non-Gaussian operations do not provide a meaningful improvement over the TMSS in practical implementations.

We now proceed to analyze the performance of QI under these engineered protocols. We present in Figs.~\ref{fig4}(c) and (d) the minimum error probability as a function of the number of $K$. We begin with Fig.~\ref{fig4}(c), where the error probability is evaluated at the values of $T$ that maximize the von Neumann entropy. In this regime, NLPA again yields the lowest error probability among all considered protocols ($T \to 1$), followed by PA ($T=0.125$), while PC and PS provide smaller improvements ($T \to 1$). Overall, all engineered protocols outperform TMSS at these optimal points~\cite{zhang2024quantum}.

In Fig.~\ref{fig4}(d), the situation changes, as the error probability is now evaluated at the values of $T$ that maximize the success probability. In this regime, none of the engineered protocols offers a genuine improvement over TMSS. The only scheme that reproduces a comparable performance is PC, which effectively mimics the behavior of TMSS. In particular, NLPA exhibits a worse performance than TMSS. 
 This demonstrates that the performance gain originates in the single-photon regime, as shown in Figs.~\ref{fig1}(c) and (d).

\begin{figure}[!htbp]
\centering
\includegraphics[width=\linewidth]{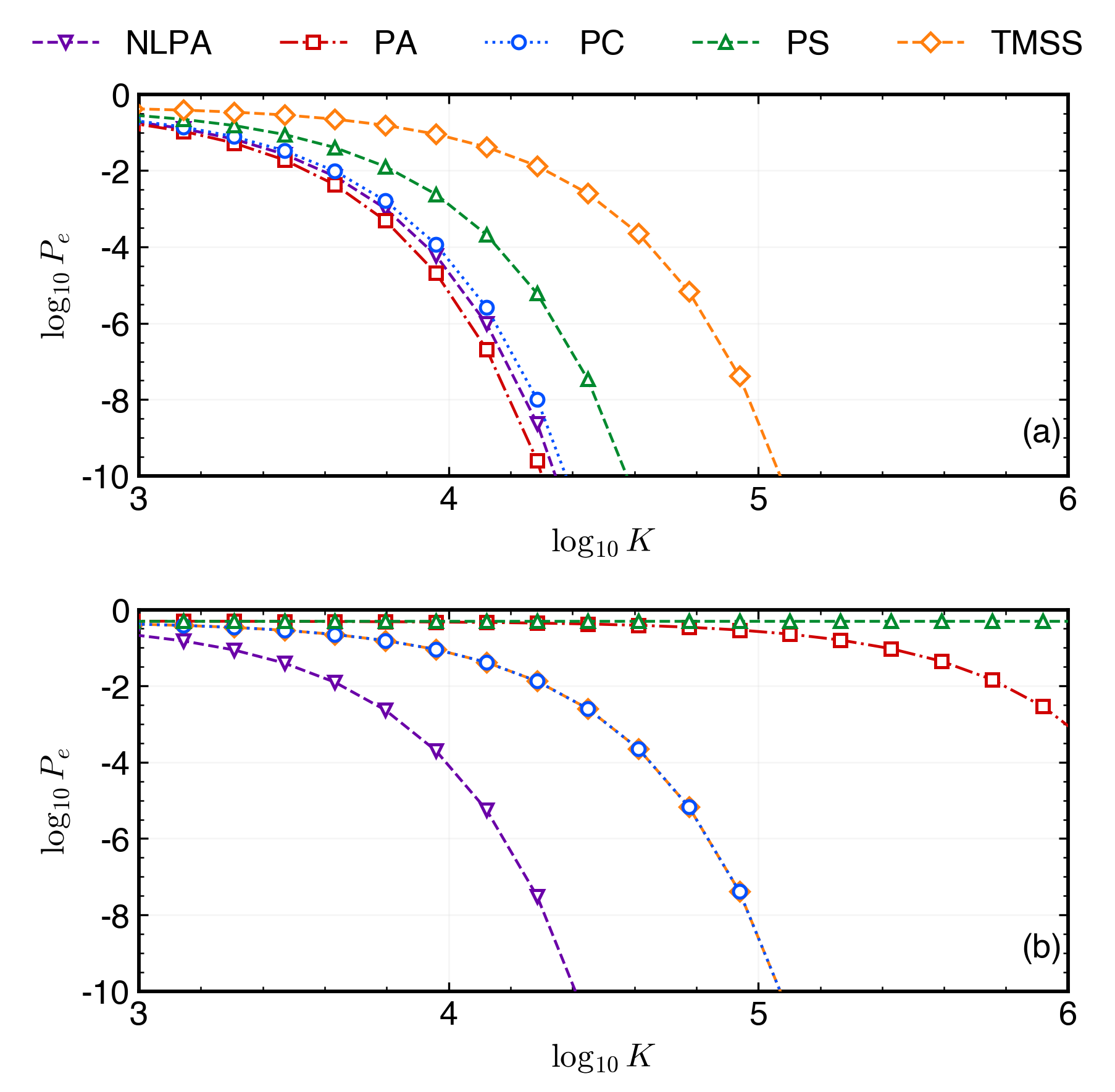}
\caption{Error probability as fuction of number of copies $K$ (a) evaluated at the transmissivity $T$ that maximizes the von Neumann entropy $E_V$ and (b) transmissivity $T$ that maximizes the success probability for non-Gaussian protocols using two auxiliary photons and the NLPA using one auxiliary photon. The TMSS case (no non-Gaussian operation) is shown as a reference. 
}
\label{fig7}
\end{figure}

As a final comparison, we contrast the NLPA implemented with a single auxiliary photon, which yields the best performance under minimal resource conditions, with local non-Gaussian operations employing two auxiliary photons, representing the most competitive higher-resource alternative. In Fig.~\ref{fig7}, we present the error probability as a function of the number of pair signal-idler $K$: panel (a) corresponds to the values of $T$ that maximize the von Neumann entropy, while panel (b) shows the results at the values of $T$ that maximize the success probability.

In Fig.~\ref{fig7}(a), the lowest error probability is achieved by the PA protocol. However, the key observation is that NLPA attains a nearly identical performance, with only a marginal difference separating the two. This is particularly significant when considering the required resources: NLPA achieves a comparable performance using a single auxiliary photon, whereas PA relies on two. Given the experimental challenges associated with generating and controlling multi-photon states, this result highlights NLPA as a more practical and scalable alternative for improving over the TMSS baseline. A markedly different behavior is observed in Fig.~\ref{fig7}(b), where the comparison is performed in the high-success-probability regime. In this case, NLPA clearly yields the lowest error probability, without any meaningful competition from the local non-Gaussian protocols. While PC reproduces a behavior similar to TMSS, PA and PS fail to provide any improvement. This demonstrates that the apparent advantage of multi-photon non-Gaussian operations in panel (a) does not translate into a realistic performance gain under experimentally relevant conditions. Altogether, these results reveal two central advantages of NLPA over the local non-Gaussian alternatives: it provides a genuine improvement in regimes of high success probability, and it does so with reduced resource requirements. This combination of performance and practicality establishes NLPA as a more viable engineered-state strategy for minimizing the error probability, thereby enhancing the discrimination task in QI.

\begin{figure}[!htbp]
\centering
\includegraphics[width=\columnwidth]{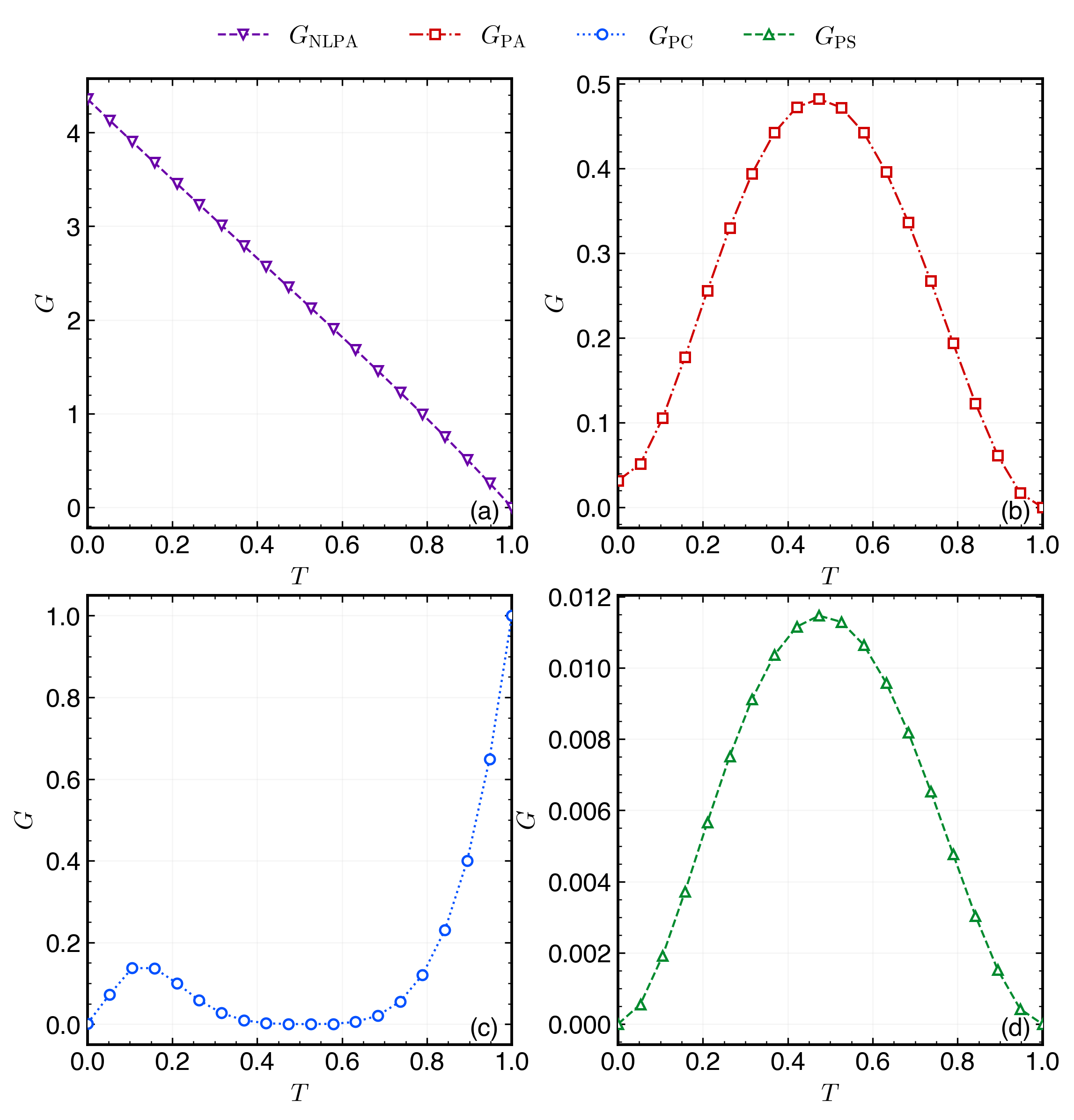}
\caption{
Error exponent ratio $G$ as a function of transmissivity $T$ for non-Gaussian protocols using two auxiliary photons and the NLPA using one auxiliary photon. Panels correspond to the ratio of (a) nonlocal photon addition (NLPA), (b) local photon addition (PA), (c) local photon catalysis (PC), and (d) local photon subtraction (PS). }
\label{fig8}
\end{figure}

In Fig.~\ref{fig8}, we present the gain parameter $G_\alpha$ as a function of the beam-splitter transmissivity $T$ for the comparison between the NLPA protocol implemented with a single auxiliary photon and the local non-Gaussian operations (PA, PC, and PS) implemented with two auxiliary photons. Figure~\ref{fig8}(a) shows that $G_{\mathrm{NLPA}}$ remains the dominant contribution, decreasing monotonically from its largest value in the low-$T$ region to zero as $T\rightarrow1$. This indicates that the single-photon NLPA retains the strongest effective exponent when the preparation probability is highest. As $T$ increases, the gain is progressively reduced because the heralding probability of the conditional protocol decreases, eventually compensating the discrimination advantage obtained from the engineered correlations. In Fig.~\ref{fig8}(b), $G_{\mathrm{PA}}$ exhibits a broad maximum near intermediate transmissivities, reaching values substantially below those obtained in the corresponding single-photon comparison. This reduction arises because, although two-photon photon-addition operations can improve the distinguishability factor $Q_\alpha$ relative to the single-photon case, the corresponding success probability is significantly lower, suppressing the overall gain. Therefore, the additional non-Gaussian resource does not translate into a larger operational advantage. The behavior of $G_{\mathrm{PC}}$ in Fig.~\ref{fig8}(c) is strongly non-monotonic. The gain remains small over most of the parameter range, with a small low-$T$ maxima, a near-vanishing minimum at intermediate transmissivities, and a rapid increase only as $T\rightarrow1$, where it approaches unity. This trend indicates that photon catalysis with two auxiliary photons becomes competitive only in a narrow high-$T$ region. However, this occurs in the high-$T$ regime, where the success probability is near maximal and the conditioned state approaches TMSS-like behavior. Therefore, a genuine non-Gaussian advantage is not observed in this case based on the  defined gain parameter.
Figure~\ref{fig8}(d) shows that $G_{\mathrm{PS}}$ remains very small throughout the full transmissivity range, displaying only a shallow maximum near $T=0.5$ and vanishing at both endpoints. This demonstrates that photon subtraction with two auxiliary photons provides the weakest effective gain among the local protocols considered. Although subtraction can modify the photon-number distribution in a favorable manner, the corresponding detection cost strongly suppresses the success-probability-weighted exponent.

These results show that increasing the number of auxiliary photons in local non-Gaussian operations can improve discrimination performance, only with the cost of the low success probabilities.
Thus, this improvement is not necessarily reflected in the gain parameter $G$. Although two-photon PA, PC, and PS schemes can yield better distinguishability, their lower preparation probabilities strongly affect the effective exponent entering $G$, thereby reducing the reported gain. In contrast, the single-photon NLPA protocol achieves a more favorable compromise between correlation enhancement and heralding efficiency. 

\subsection{Performance Under Photon Loss}
\label{sec3e}

In QI, as in many open quantum-system scenarios, the signal mode inevitably experiences photon loss during its propagation through the sensing channel. In realistic implementations, such losses are unavoidable and constitute one of the primary mechanisms degrading the performance of quantum protocols. In particular, photon loss  reduces the strength of the returned signal and weakens the useful correlations between the signal and the stored idler.

Photon loss is commonly modeled as occurring after the signal mode interacts with a weakly reflecting target, whereby the returning mode $B'$ undergoes additional attenuation during propagation through the detection channel. This loss process can be described by a beam splitter interaction between the returning mode and an environmental vacuum mode $D$. The corresponding unitary transformation is written as $\hat V_{\mathrm{BS}}=e^{\phi F}$, where $\phi=\arctan\!\left(\sqrt{\frac{1-\eta}{\eta}}\right)$ and
$
F=\hat a_{B'}\hat a_{D}^{\dagger}-\hat a_{B'}^{\dagger}\hat a_{D}$.
Here $\hat a_{B'}$ and $\hat a_{D}$ denote the annihilation operators associated with the returning mode and the vacuum environment, respectively, while $\eta$ represents the transmissivity of the lossy channel~\cite{zhang2014quantum}. The state under hypothesis $H_1$, including propagation through the loss channel, is therefore given by
\begin{equation}
\rho'_{1}
= \Tr_{D} \!\left[
\hat V_{\mathrm{BS}}(\eta)
\big(
\rho_{1} \otimes \ket{0}\bra{0}_{D}
\big)
\hat V_{\mathrm{BS}}^{\dagger}
\right].
\end{equation}
Here $\rho_1$ is defined by Eq.~\eqref{rho1}. The loss channel only affects the returned signal under the target-present hypothesis. In contrast, when the target is absent ($H_0$), the receiver collects only environmental thermal noise, which is already defined at the detection stage and therefore does not undergo additional attenuation in this model. As a result, the state $\rho_0$ remains unchanged. Consequently, the discrimination problem in the QI  reduces to determining the minimum error probability for distinguishing between the states $\rho_0$ and the noisy returned state $\rho'_1$.

\begin{figure}[!htbp]
\centering
\includegraphics[width=\linewidth]{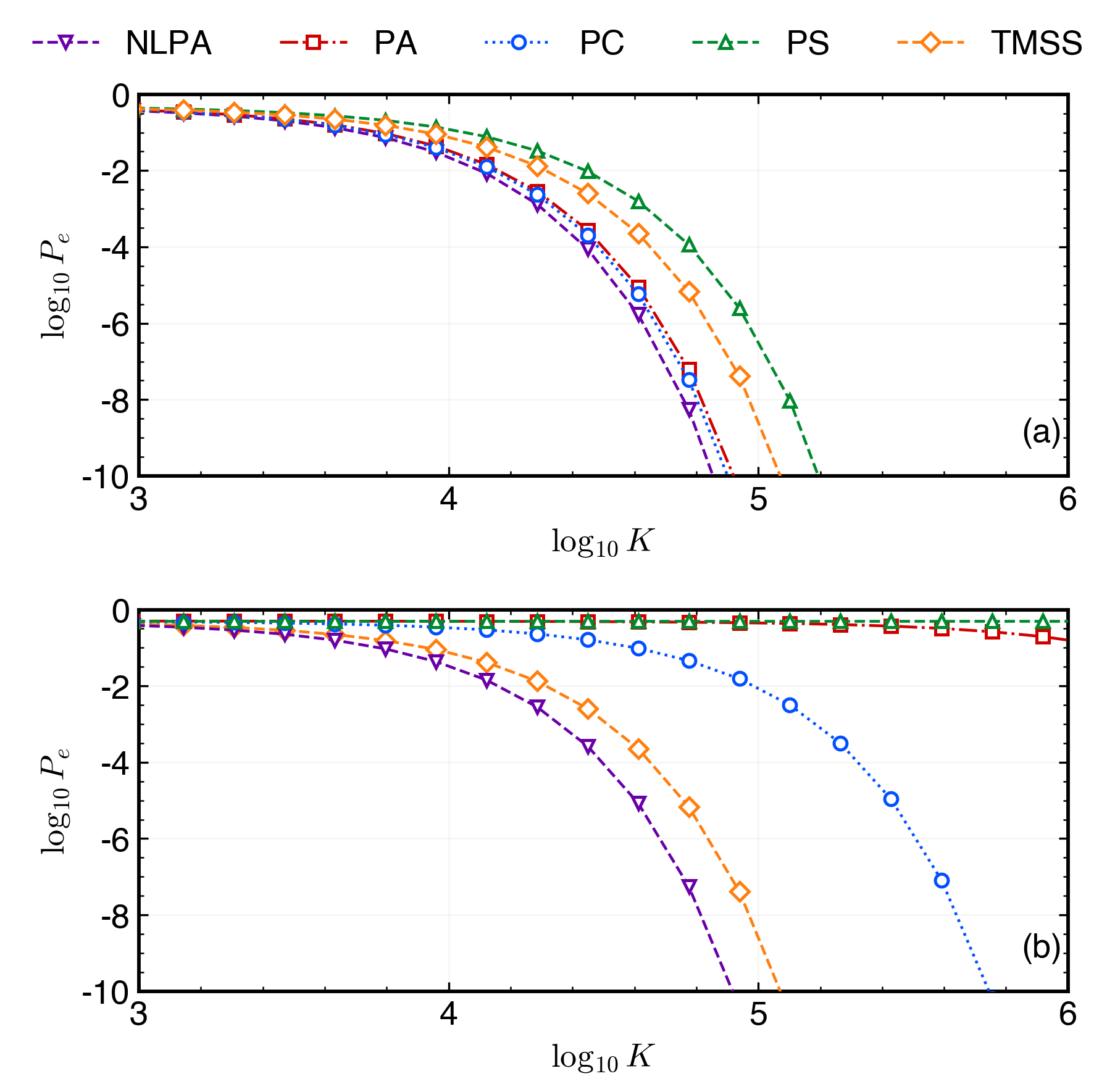}
\caption{
Error probability as fuction of number of copies $K$ (a) evaluated at the transmissivity $T$ that maximizes the von Neumann entropy $E_V$ and (b) transmissivity $T$ that maximizes the success probability for non-Gaussian protocols using two auxiliary photons and the NLPA using one auxiliary photon. The noiseless TMSS case is shown as a reference. All states are shown after propagation through a photon loss channel with loss parameter $\eta = 0.1$.
}
\label{fig11}
\end{figure}

In this context, Fig.~\ref{fig11}(a) shows the performance of the local non-Gaussian operations (PS, PA, and PC) implemented with two auxiliary photons, together with the single-photon NLPA protocol, using the noiseless TMSS case as a reference, the photon-loss channel significantly degrades the performance of all local non-Gaussian operations compared to the noiseless case in Fig.~\ref{fig7}(a). In contrast to the ideal scenario, NLPA now achieves the lowest error probability, while PA, previously the best-performing protocol, exhibits a marked deterioration. This behavior indicates that the state generated by NLPA in the high-von Neumann entropy regime is more resilient to loss, as its dominant components preserve a structure that remains effective for discrimination after channel attenuation. Consequently, NLPA consistently outperforms all other protocols, establishing it as the most robust strategy under realistic conditions. The most pronounced separation is observed in Fig.~\ref{fig11}(b), where NLPA clearly dominates in terms of lowest error probability. In this regime, PS and PA exhibit significantly poorer performance, while PC effectively reproduces the behavior of TMSS without providing any improvement. This result is particularly relevant because it occurs in the high-success-probability regime, where the engineered state is dominated by the $\ket{\phi_0}$ component [Eq.~\eqref{phin}]. The Bell-like structure of this state enables a more stable response under photon loss, preserving distinguishability between the hypotheses despite attenuation. Consequently, NLPA maintains a clear advantage in realistic noisy conditions, combining robustness to loss with improved discrimination performance.

\begin{figure}[!htbp]
\centering
\includegraphics[width=\columnwidth]{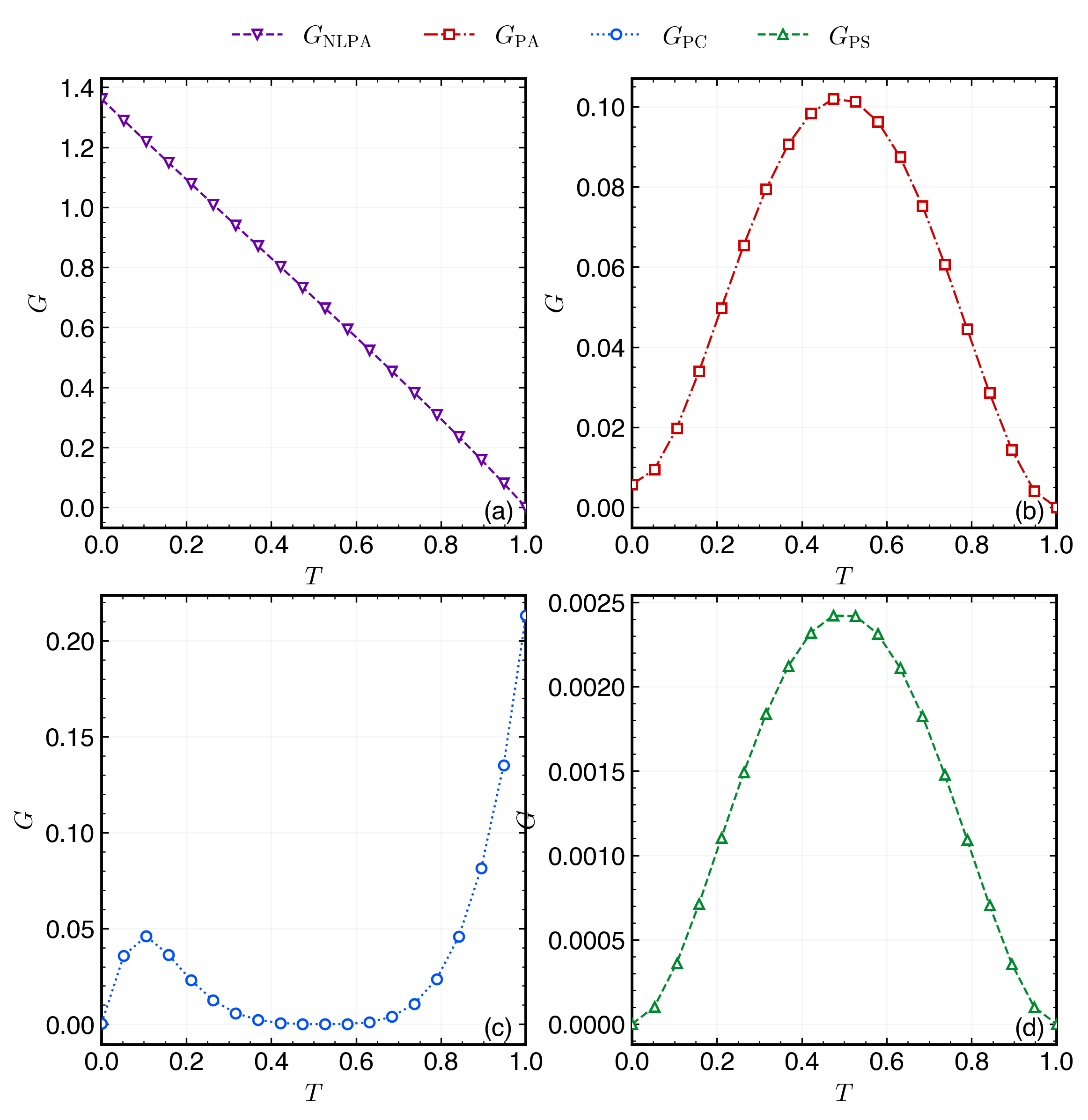}
\caption{
Error exponent ratio $G$ as a function of transmissivity $T$ for non-Gaussian protocols using two auxiliary photons and the NLPA using one auxiliary photon. Panels correspond to the ratio of (a) nonlocal photon addition (NLPA), (b) local photon addition (PA), (c) local photon catalysis (PC), and (d) local photon subtraction (PS).  All states are shown after propagation through a photon loss channel with loss parameter $\eta = 0.1$. }
\label{fig12}
\end{figure}

Next, we examine the gain parameter $G$ in the presence of a photon-loss channel, as shown in Fig.~\ref{fig12}, for the comparison between the single-photon NLPA protocol and the local non-Gaussian operations implemented with two auxiliary photons. Figure~\ref{fig12}(a) shows that $G_{\mathrm{NLPA}}$ preserves the same monotonic decrease with $T$ observed in the absence of loss, but with a substantially reduced magnitude. Its maximum value in the low-$T$ region drops from the lossless case, and the gain vanishes as $T\rightarrow1$. This behavior shows that photon loss attenuates the advantage of NLPA, yet still leaves it as the only protocol retaining an appreciable gain over a broad transmissivity range. Then, this indicates that the dominant components of the NLPA transmitter remain comparatively more resilient to attenuation, so that the effective discrimination capability survives even after the loss channel. In Fig.~\ref{fig12}(b), $G_{\mathrm{PA}}$ retains a broad maximum near intermediate transmissivities, but its amplitude is strongly suppressed compared with the corresponding lossless case. Thus, although two-photon PA can still improve the conditional-state distinguishability, the combined effect of photon loss and low preparation probability strongly reduces the effective exponent. The behavior of $G_{\mathrm{PC}}$ in Fig.~\ref{fig12}(c) remains non-monotonic, with a shallow low-$T$ structure, a near-vanishing intermediate region, and a recovery only as $T\to1$. However, this recovery is much weaker than in the lossless case, and the gain remains small throughout most of the parameter space. This indicates that photon loss strongly suppresses the already limited operational benefit of the two-photon PC. Even when the conditioned state becomes more favorable, the gain remains constrained by the reduced effective exponent. Figure~\ref{fig12}(d) shows that $G_{\mathrm{PS}}$ is the most strongly suppressed of all cases. The curve remains close to zero over the full range of $T$, with only a very small maximum around intermediate $T$. Thus, two-photon PS is particularly fragile under attenuation. Any improvement at the level of conditional-state discrimination is almost entirely offset by the combined effects from loss and low generation probability.

These results show that photon loss suppresses the gain parameter for all protocols, with the strongest reduction occurring for the local two-photon non-Gaussian schemes. By contrast, the single-photon NLPA protocol is the least degraded and the only scheme that retains a modest gain under loss. 

\subsection{Signal-to-Noise Ratio and Detection Scheme}
\label{sec3f}

The performance of the QI under a given receiver scheme is characterized by the signal-to-noise ratio (SNR). In this framework, an observable $\hat{O}$ constructed from the return mode and the idler modes is measured. The corresponding mean values under the two hypotheses $H_0$ (target absent) and $H_1$ (target present) are defined as
\begin{equation}
M_j = \langle \hat{O} \rangle_{H_j}, \label{means}
\end{equation}
with associated variances
\begin{equation}
\Delta M_j = \langle \hat{O}^2 \rangle_{H_j} - \langle \hat{O} \rangle_{H_j}^2, \label{variances}
\end{equation}
for $j \in \{0,1\}$.

A threshold decision rule is applied to the measurement outcome $s$ obtained from $K$ independent signal-idler mode pairs. Given a threshold $M_{\mathrm{th}}$, we decide $H_0$ if $s > M_{\mathrm{th}}$ and $H_1$ otherwise. Assuming equal prior probabilities, the resulting error probability is
\begin{equation}
P_e
=
\frac{1}{2}
\left[
P(1|H_0) + P(0|H_1)
\right],
\end{equation}
where $P(1|H_0)$ and $P(0|H_1)$ correspond to the false-alarm and miss-detection probabilities, respectively.

For a large number of mode pairs ($K \gg 1$), the measurement outcome $s$ approaches a Gaussian distribution by the central limit theorem, with mean $K M_j$ and variance $K \Delta M_j$~\cite{guha2009gaussian,jo2021quantum}. The conditional error probabilities are then given by ~\cite{guha2009gaussian,jo2021quantum}
\begin{subequations}
\begin{align}
P(1|H_0) &= \frac{1}{2}\operatorname{erfc}\!\left(
\frac{K M_0 - M_{\mathrm{th}}}{\sqrt{2K \Delta M_0}}
\right),\\
P(0|H_1) &= \frac{1}{2}\operatorname{erfc}\!\left(
\frac{M_{\mathrm{th}} - K M_1}{\sqrt{2K \Delta M_1}}
\right).
\end{align}
\end{subequations}
The threshold that minimizes the error probability is obtained by symmetrizing the two error contributions, yielding
\begin{equation}
M_{\mathrm{th}}
=
K\,\frac{\sqrt{\Delta M_1}\,M_0 + \sqrt{\Delta M_0}\,M_1}
{\sqrt{\Delta M_0} + \sqrt{\Delta M_1}}.
\end{equation}

Substituting this optimal threshold, the minimum error probability becomes
\begin{equation}
P_e
=
\frac{1}{2}\operatorname{erfc}\!\left[
\frac{\sqrt{K}\,(M_0 - M_1)}
{\sqrt{2}\left(\sqrt{\Delta M_0} + \sqrt{\Delta M_1}\right)}
\right].
\end{equation}

This naturally defines the SNR as
\begin{equation}
\mathrm{SNR}
=
\frac{K\,(M_0 - M_1)^2}
{2\left(\sqrt{\Delta M_0} + \sqrt{\Delta M_1}\right)^2},
\label{SNR}
\end{equation}
and, in decibel units, $\mathrm{SNR}_{\mathrm{dB}} = 10 \log_{10}(\mathrm{SNR})$.
Several receiver strategies have been investigated and proposed. Among the various receiver strategies, double homodyne detection (dHD) has emerged as an effective and experimentally accessible scheme, exhibiting strong performance even in noisy environments. In contrast to receivers based on nonlinear interactions, the dHD scheme relies solely on linear optics, enabling a simpler and more feasible implementation. For Gaussian probe states with vanishing mean, discrimination is governed by second-order moments, such that the measurement directly accesses the phase-sensitive inter-mode correlations encoded in the covariance matrix, which underpin the performance advantage in QI~\cite{jo2021quantum}. The corresponding observable is given by
\begin{equation}
\hat{O}_{\mathrm{dHD}}
=
\hat{a}_A^\dagger \hat{a}_A
-
\left(
\hat{a}_B^\dagger \hat{a}_A^\dagger
+
\hat{a}_B \hat{a}_A
\right)
+
\hat{a}_B \hat{a}_B^\dagger,
\end{equation}
which arises from mixing the return and idler modes on a $50{:}50$ beam splitter followed by homodyne measurements on the output ports. Given the favorable performance of this receiver, several local non-Gaussian operations such as PC, PA, and PS have been investigated within this measurement scheme. Double photon addition provides the largest enhancement relative to the TMSS, with improvements of approximately $9\,\mathrm{dB}$, while single-photon subtraction yields a more modest gain of about $6\,\mathrm{dB}$~\cite{zhang2024quantum}. 

\begin{figure}[!htbp]
\centering
\includegraphics[width=\linewidth]{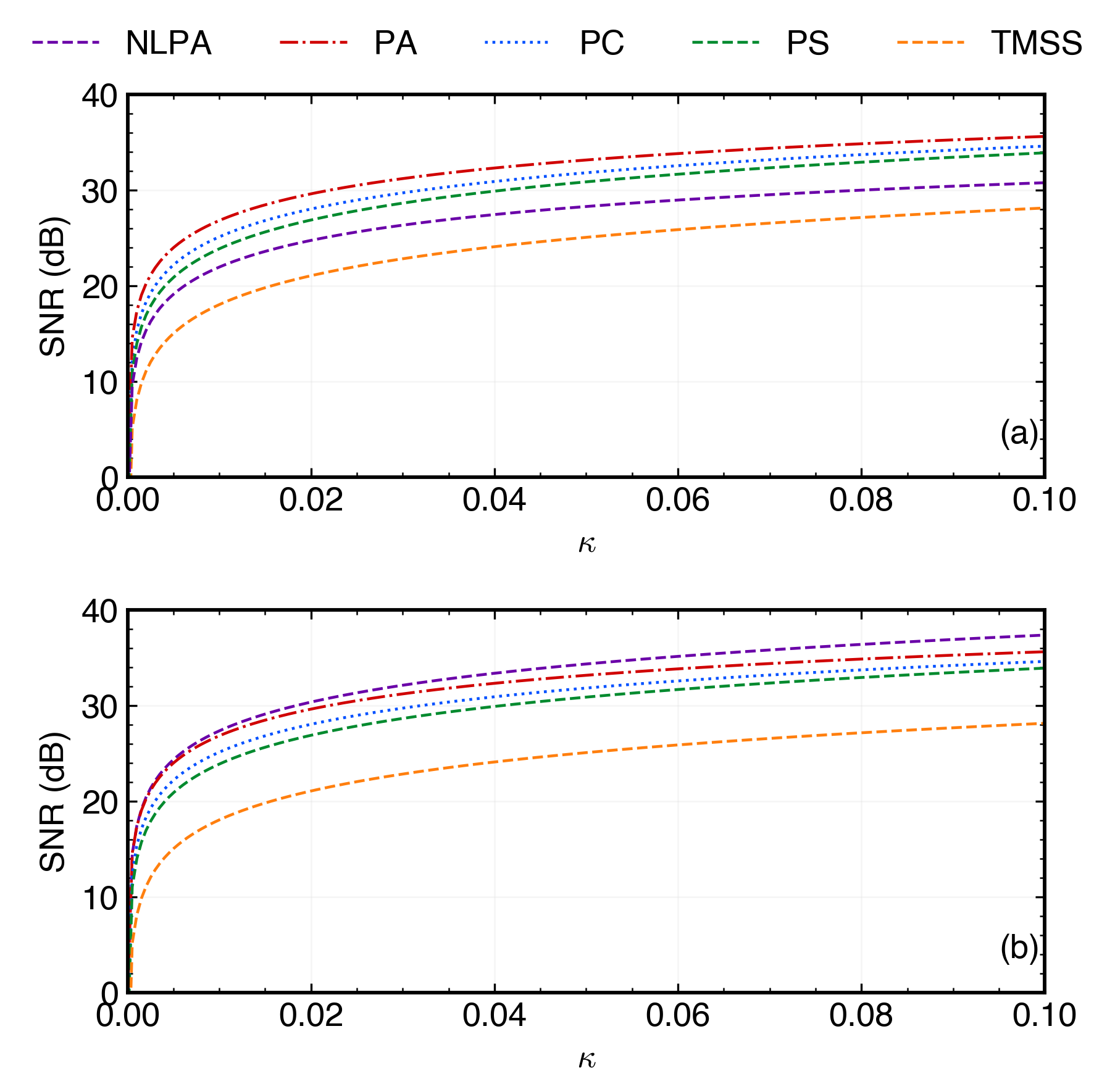}
\caption{The SNR as a function of the target reflectivity $\kappa$. (a) All evaluated using double homodyne detection. (b) double homodyne detection for TMSS and local non-Gaussian operation, while for the NLPA is evaluated using the difference photon count number after the beam splitter scheme.}
\label{fig9}
\end{figure}

Fig.~\ref{fig9}(a) shows the SNR using dHD as a function of the target reflectivity $\kappa$ for the three local non-Gaussian operations (PS, PA, and PC) implemented with two auxiliary photons, together with NLPA employing a single auxiliary photon, as these represent the most competitive configurations identified in this work, evaluated at $K = 10^6$ signal-idler pairs and at their respective optimal entanglement $T$. As expected, the SNR increases monotonically with $\kappa$, since larger reflectivity enhances the returned signal and improves the distinguishability between the target-present and target-absent hypotheses. Among the local non-Gaussian operations, PA and PC achieve the highest SNR across most of the explored range, while PS also improves over the Gaussian TMSS baseline but to a lesser extent. However, the NLPA transmitter does not yield the largest SNR under this receiver configuration; although it outperforms the TMSS baseline, it remains below PA, PC, and PS. This behavior is consistent with the fact that the dHD-based SNR is governed by phase-sensitive correlations $\langle \hat a_B^\dagger \hat a_A^\dagger \rangle$ and $\langle \hat a_B \hat a_A \rangle$, which are suppressed by the Bell-type structure of Eq.~\eqref{phin}, while the conditioning process enhances higher-order fluctuations, increasing the variances $V_0$ and $V_1$ and thus reducing the SNR. Moreover, the non-Gaussian coherence structure generated by the conditional preparation is not optimally matched to the dHD measurement operator $\hat{O}_{\mathrm{dHD}}$, so part of the available nonclassical resource is not fully captured by this detection scheme.

Motivated by the limitations of the dHD receiver and by the fact that, in the weak-squeezing regime, we consider a measurement strategy tailored to exploit the exchange-type correlations $\langle \hat a_A^\dagger \hat a_B \rangle$ and $\langle \hat a_A \hat a_B^\dagger \rangle$. This naturally leads to a detection scheme based on interference at a $50{:}50$ beam splitter followed by photon counting, where the photon-number difference directly probes these coherences~[see Fig.~\ref{sch1}(c)]. The beam splitter transformation is described by the unitary operator
\begin{equation}
U_{\mathrm{BS}} = \exp\!\left[\frac{\pi}{4}\left(\hat a_A \hat a_B^\dagger - \hat a_A^\dagger \hat a_B\right)\right], \label{BS}
\end{equation}
which mixes the idler and return modes and converts their mutual coherence into a measurable photon-number imbalance. Under this transformation, the input modes $(A,B)$ are mapped to the output modes $(1,2)$ as
\begin{subequations}
\begin{align}
\hat a_1 &= \frac{1}{\sqrt{2}}\left(\hat a_A + \hat a_B\right),\\
\hat a_2 &= \frac{1}{\sqrt{2}}\left(\hat a_A - \hat a_B\right),
\end{align}
\end{subequations}
and the photon-number difference operator
\begin{equation}
\hat O = \hat n_1 - \hat n_2
\end{equation}
can be expressed in terms of the input modes as
\begin{equation}
\hat O= \hat a_A^\dagger \hat a_B + \hat a_B^\dagger \hat a_A. \label{eqOprop}
\end{equation}
This approach is consistent with recent results showing that Bell-like non-Gaussian states combined with simple interference-based detection can outperform Gaussian TMSV-based protocols under comparable conditions~\cite{wang2024detection}.

Fig.~\ref{fig9}(b) shows the performance of NLPA under the proposed detection scheme, compared with the local non-Gaussian operations PA, PS, and PC, as well as the TMSS baseline under balanced dHD detection. All results are evaluated at $K = 10^6$ and at the respective optimal values of the beam-splitter transmissivity $T$ for each protocol. The NLPA outperforms the TMSS by approximately $10\,\mathrm{dB}$ and surpasses PA, the best-performing non-Gaussian transmitter under homodyne detection. This enhancement originates from the structure of the optimized observable in Eq.~\eqref{eqOprop}, which exploits the correlation redistribution in Eq.~\eqref{phin}, where the state acquires a Bell-like form with dominant contributions from $\langle \hat a_A^\dagger \hat a_B \rangle$ and $\langle \hat a_A \hat a_B^\dagger \rangle$. These results indicate that the NLPA is well matched to receiver strategies that exploit first-order inter-mode coherences, which are maximized in this protocol, enabling significantly enhanced SNR compared with the TMSS and other local non-Gaussian protocols. This performance is achieved while retaining practical advantages, including the use of a single auxiliary photon, a high success probability, and a broad range of parameter $T$ over which high entanglement is maintained.

\section{Conclusion}
\label{sec4}

In this work, we investigated quantum illumination using an engineered entangled probe generated via the nonlocal photon addition protocol and compare it with standard local non-Gaussian operations, including photon addition, subtraction, and catalysis. Using the error probability as the performance metric, we showed that the nonlocal non-Gaussian photon addition protocol provides a significant advantage under realistic experimental conditions, achieving improved performance and higher success probability with comparable resources, even in the presence of photon loss. In particular we demonstrated that nonlocal single-photon addition outperforms other scenarios, providing an optimal protocol with minimal resource for quantum illumination applications.

Furthermore, using a feasible receiver based on a $50{:}50$ beam splitter and photon-number difference measurement, we demonstrate a significant enhancement in the signal-to-noise ratio relative to the TMSS and competing local non-Gaussian protocols, achieved with a single auxiliary photon. These results establish the NLPA as a competitive and experimentally viable approach for QI in practically relevant regimes.

\bibliography{refs}

\end{document}